# The performance of univariate goodness-of-fit tests for normality based on the empirical characteristic function in large samples

By J. M. VAN ZYL

Department of Mathematical Statistics and Actuarial Science, University of the
Free State, Bloemfontein, South Africa

wwjvz@ufs.ac.za

SUMMARY

An empirical power comparison is made between two tests based on the empirical characteristic function and some of the best performing tests for normality. A simple normality test based on the empirical characteristic function calculated in a single point is shown to outperform the more complicated Epps-Pulley test and the frequentist tests included in the study in large samples.

*Key words:* Normality test; Empirical characteristic function; Cumulant; Goodness-of-fit

1. INTRODUCTION

Several goodness-of-fit tests based on the empirical characteristic function (ecf) are available. Feuerverger and Mureika (1977) developed a test for symmetry and this was extended by Epps and Pulley (1983) to test univariate normality. Henze and Baringhaus (1988) extended the Epps-Pulley test to test multivariate normality and this is called the BHEP test. A review paper with comments of procedures based on the ecf is the paper by Meintanis (2016) and also the book by Ushakov (1999). The Epps-Pulley approach is still the main approach for testing normality of tests based on the ecf and is based on an integral over the weighted squared distances between the ecf and the expected characteristic function. A weakness of the test is that the asymptotic null-distribution is not very accurate and otherwise intractable and complicated in finite samples (Taufer (2016), Swanepoel and Alisson (2016)).

In this work a test is proposed and asymptotic distributional results derived using the work by Murota and Takeuchi (1981). They derived a location and scale invariant test using studentized observations and showed that the use of a single value when calculating the ecf is sufficient to get good results with respect to power when testing hypotheses. Csörgő (1986) derived a multivariate extension of the asymptotic results derived by Murota and Takeuchi (1981).

A simulation study is conducted to compare the power of the proposed test against the Epps-Pulley test and five of the most recognized goodness-of-fit tests for normality. The proposed test with asymptotic properties similar to those of the test of Murota and Takeuchi (1981) performs reasonably in small sample, but excellent in large samples with respect to power. The test statistic is a simple normal test which will perform better as the sample increases and it was found to dominate the much more complicated Epps-Pulley test.

Murota and Takeuchi (1981) compared their test against a test based on the sample kurtosis and conducted a small simulation study. Various overview simulation studies were conducted to investigate the performance of tests for normality. One of the most cited papers is the work by Yap and Sim (2011), but they did not include a goodness-of-fit test based on the empirical characteristic function. A paper which included a very large selection of tests for normality is the work by Romao et al. (2013), but the test of Murota and Takeuchi (1981) was not included in this study.

The tests included are the Jarque–Bera, Shapiro-Wilk, Lilliefors, Anderson-Darling and D'Agostino and Pearson tests. The focus will be on unimodal symmetric distributions and large sample sizes, that is sample sizes larger than fifty.

Murota and Takeuchi (1981) proved that the square of the modulus of the empirical characteristic function converges weakly to a complex Gaussian process where the observations are standardized using affine invariant estimators of location and scale and they derived an expression for the asymptotic variance. Let $X_1,...,X_n$ be an i.i.d. sample of size $n$, from a distribution $F$. The characteristic function is $E(e^{itX}) = \phi(t)$ and it is estimated by the ecf

$$\hat{\phi}_F(t) = \frac{1}{n}\sum_{j=1}^{n} e^{itX_j}, \qquad (1)$$

The studentized sample is $Z_1,...,Z_n$, where $Z_j = (X_j - \hat{\mu}_n)/\hat{\sigma}_n$, $j=1,...,n$, with $\hat{\mu}_n = \bar{X}_n$ and $\hat{\sigma}_n^2 = S_n^2$. Denote the ecf, based on the studentized data by $\hat{\phi}_S(t) = (1/n)\sum_{j=1}^n e^{itZ_j}$. The statistic proposed to test normality is

$$v_n(1) = \log(|\hat{\phi}_S(1)/\exp(-1/2)|), \qquad (2)$$

where $\sqrt{n}\,(v_n(1)) \sim N(0, 0.0431)$ asymptotically. Absolute value denotes the modulus of a complex number if the argument is complex.

The expression is

$$I_n = \int_{-\infty}^{\infty} |\hat{\phi}_S(t) - \hat{\phi}_0(t)|^2 \, w(t)dt,$$

with $\phi_0(t)$ denoting the ecf of a standard normal, $w(t)$ a weight function which is of the same form as a normal density with mean zero and variance the sample estimate of the variance. Of the many variations using the ecf to test goodness-of-fit tests, this expression attracted the most interest and is still used Meintanis (2016). Epps and Pulley (1983) used this expression and derived a test for normality using a weight function which has the form of a standard normal density. They gave an exact expression for the characteristic function of the normal distribution. Henze (1990) derived a large sample approximation for this test and used Pearson curves to approximate the distribution. It is shown in the simulation study that the proposed test with similar properties as that of Murota and Takeuchi (1981) and calculated in a single point without using a weight outperforms the Epps-Pulley test in the cases considered with respect to power.

## 2. MOTIVATION AND ASYMPTOTIC VARIANCE OF THE TEST STATISTIC

A motivation will be given in terms of the cumulant generating function. The normal distribution has the unique property that the cumulant generating function cannot be a finite-order polynomial of degree larger than two, and the normal distribution is the only distribution for which all cumulants of order larger than 3 are zero (Cramér (1946), Lukacs, (1972)).

The motivation for the test will be shown by using the moment generating function, but experimentation showed that the use of the characteristic function rather than the moment generating function gives much better results when used to test for normality. Consider a random variable $X$ with distribution $F$, mean $\mu$ and variance $\sigma^2$. The cumulant generating function $\mathrm{K}_F(t)$ of $F$ can be written as $\mathrm{K}_F(t) = \sum_{r=1}^{\infty} \kappa_r t^r / r!$, where $\kappa_r$ is the r-th cumulant. The first two cumulants are $\kappa_1 = E(X) = \mu$, $\kappa_2 = Var(X) = \sigma^2$. Since $\mathrm{K}_F(t)$ is the logarithm of the moment generating function, the moment generating function can be written as $M_F(t) = E(e^{tX}) = e^{\mathrm{K}_F(t)}$. It follows that

$$\mathrm{K}_F(t) = \log(E(e^{tX}))$$

$$= \sum_{r=1}^{\infty} \kappa_r t^r / r!$$

$$= \left( \kappa_1 t + \kappa_2 t^2 / 2 \right) + \left( \sum_{r=3}^{\infty} \kappa_r t^r / r! \right)$$

$$= \left( \mu t + \sigma^2 t^2 / 2 \right) + \left( \sum_{r=3}^{\infty} \kappa_r t^r / r! \right).$$

Let $F_N$ denote a normal distribution with a mean $\mu$ and variance $\sigma^2$. $M_N(t)$ denotes the moment generating function of the normal distribution with cumulant generating function $\mathrm{K}_N(t) = \mu t + \tfrac{1}{2} \sigma^2 t^2$. The logarithm of the ratio of the moment generating functions of $F$ and $F_N$ is given by

$$\log(M_F(t) / M_N(t)) = \log(\exp(\mathrm{K}_F(t) - \mathrm{K}_N(t)))$$

$$= [(\mu t + \tfrac{1}{2} \sigma^2 t^2) + \sum_{r=3}^{\infty} \kappa_r t^r / r!] - (\mu t + \tfrac{1}{2} \sigma^2 t^2)$$

$$= \mathrm{K}_N(t) + \sum_{r=3}^{\infty} \kappa_r t^r / r! - \mathrm{K}_N(t))$$

$$= \sum_{r=3}^{\infty} \kappa_r t^r / r!. \qquad (3)$$

If $F$ is a normal distribution, the sum given in (1) is zero. Replacing $\mathrm{K}_F(t) - \mathrm{K}_N(t)$ by $\mathrm{K}_F(it) - \mathrm{K}_N(it)$ it follows that

$$\log(|\log(\phi_F(t)) - \log(\phi_N(t))|) = \log(|\mathrm{K}_F(it) - \mathrm{K}_N(it)|)$$

$$= \sum_{r=0}^{\infty} \kappa_{4+2r} t^{4+2r} / (4+2r)!,$$

which would be equal to zero when the distribution *F* is a normal distribution and this expression can be used to test for normality. Murota and Takeuchi (1981) used the fact that the square root of the log of the modulus of the characteristic function of a normal distribution is linear in terms of $t$, in other words $(-\log(|\phi_N(t)|^2))^{1/2}$ is a linear function of $t$.

An asymptotic variance of $v_n(t) = \log(|\hat{\phi}_{NS}(t)/\exp(-t^2/2)|)$ can be found by using the delta method and the results of Murota and Takeuchi (1981). Let $\hat{\phi}_{NS}(t)$ denote the ecf calculated in the point $t$ using studentized normally distributed observations. They showed that the process defined by

$$\tilde{Z}(t) = \sqrt{n}(|\hat{\phi}_{NS}(t)|^2 - \exp(-t^2)), \qquad (4)$$

converges weakly to a zero mean Gaussian process and variance

$$E(\tilde{Z}^2(t)) = 4\exp(-2t^2)(\cosh(t^2) - 1 - t^4/2). \qquad (5)$$

Note that $\hat{\phi}_{NS}(t) = e^{-t^2/2}$ and by applying the delta method it follows that

$$Var(|\hat{\phi}_{NS}(t)|^2) \approx Var(|\hat{\phi}_{NS}(t)|)(2|\hat{\phi}_{NS}(t)|)^2,$$

thus $Var(|\hat{\phi}_{NS}(t)|) \approx Var(|\hat{\phi}_{NS}(t)|^2)/(2|\hat{\phi}_{NS}(t)|)^2$.

By applying the delta method again it follows that

$$Var(v(t)) = Var(\log(|\hat{\phi}_{NS}(t)/e^{-1/2}|))$$

$$\approx (1/|\hat{\phi}_{NS}(t)|^2)Var(|\hat{\phi}_{NS}(t)|)$$

$$= Var(|\hat{\phi}_{NS}(t)|^2)/4(|\hat{\phi}_{NS}(t)|)^4$$

$$= (\cosh(t^2) - 1 - t^4/2)/n. \tag{6}$$

The statistic $v_n(t)$ converges weakly to a Gaussian distribution with mean zero and variance $Var(v_n(t)) = (\cosh(t^2) - 1 - t^4/2)/n$, where $\sqrt{n}\,(v_n(1)) \sim N(0, 0.0431)$ asymptotically.

$$Var(v_n(1)) = 0.0431/n, \quad t = 1. \tag{7}$$

Reject normality if

$$|v_n(1)/(\sqrt{0.0431/n}| = |4.8168\sqrt{n}v_n(1)| > z_{1-\alpha/2}. \tag{8}$$

In the following figure the average of the log the modulus calculated in the point, $t = 1$, using standard normally distributed samples, for various samples sizes is shown. The solid line is where studentized observations were used and the dashed line where the ecf is calculated using the original sample. It can be seen that there is a large bias in small samples and the studentized ecf has less variation.

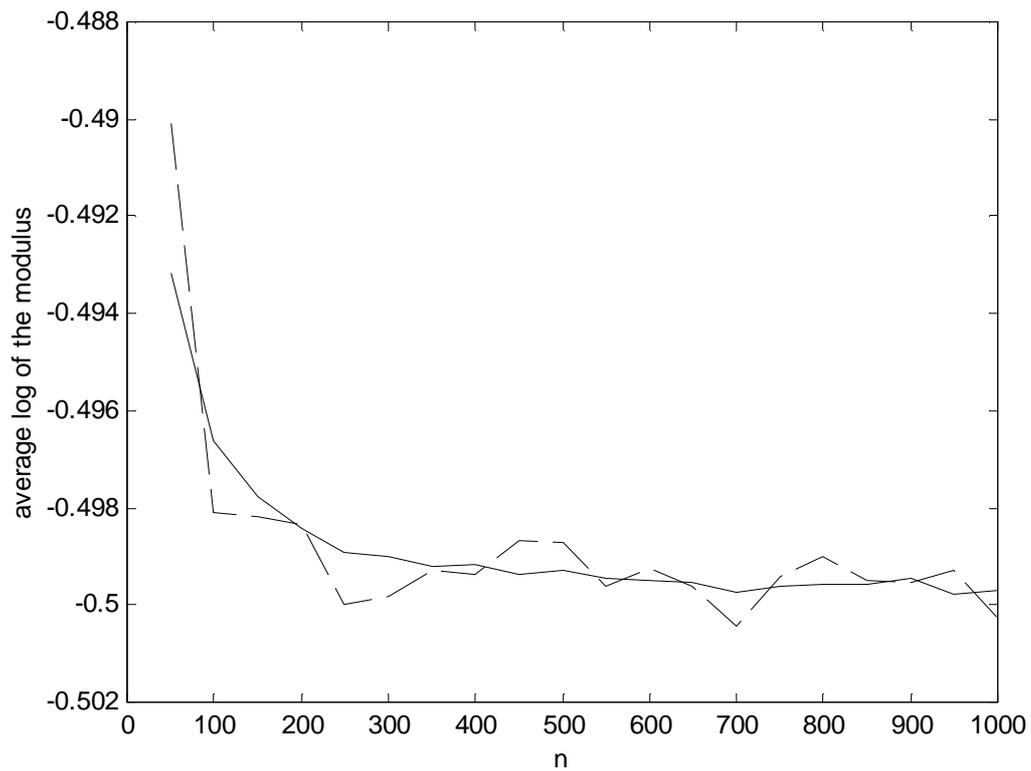

Fig. 1. Plot of the average log of the modulus of the ecf for various sample sizes calculated using $m = 5000$ calculated using samples form a standard normal distribution. The solid line is where studentized observations are used and the dashed line using the original sample. Calculated in the point $t=1$, and the expected value is -0.5.

In the following histogram 5000 simulated values of $v_n(1)/(\sqrt{0.0431/n}\,|=4.8168\sqrt{n}v_n(1)$ are shown, where the $v_n(1)'s$ are calculated using simulated samples of size $n=1000$ from a standard normal distribution.

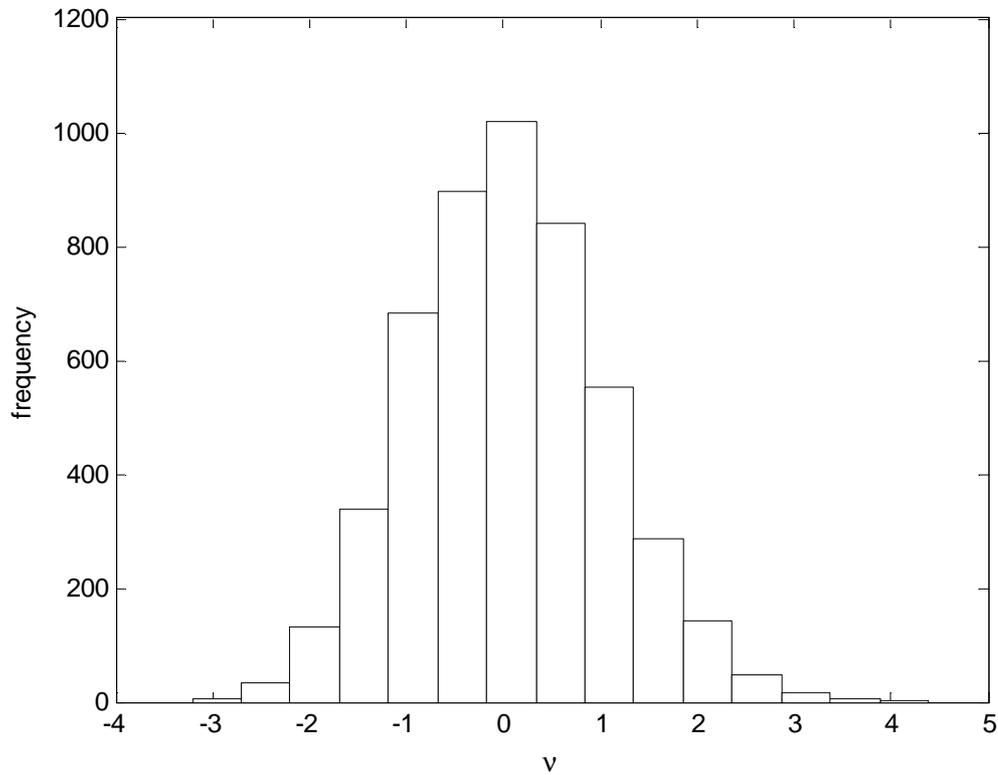

Fig. 2. Histogram of $m=5000$ simulated values of $v_n(1)/(\sqrt{0.0431/n})$, with $n=1000$. Calculated using normally distributed samples, data standardized using estimated parameters.

In Figure 3 the variance of $v(1)$ is estimated for various sample sizes, based on 1000 estimated values of $v(1)$ for each sample size considered. The estimated variances is plotted against the asymptotic variance $Var(v_n) = 0.0431/n$.

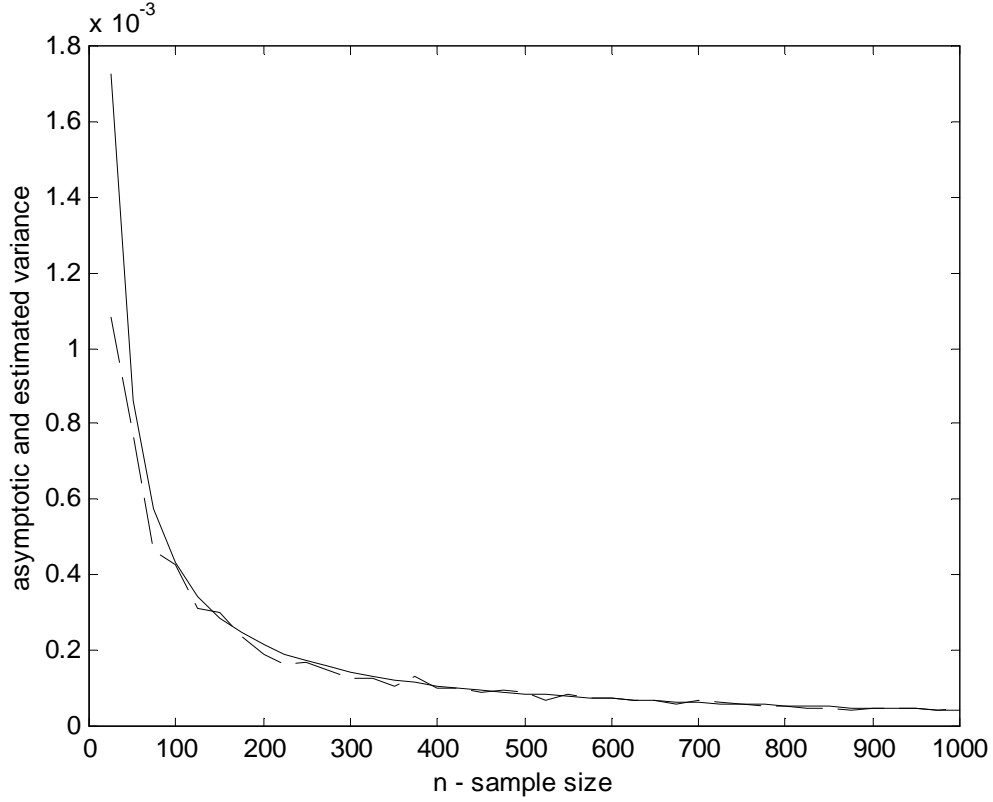

Fig. 3. Estimated variance of $v_n$ and asymptotic variance for various values of *n*. Dashed line the estimated variance. Estimated variance calculated using 1000 normally distributed samples, data standardized using estimated parameters.

There are a few variations of the parameters used when choosing the weight function for the Epps-Pulley test, but a version suggested by Epps and Pulley to be used as an omnibus test is when choosing the weigh function a normal density with mean zero and variance the sample estimate of the variance, that is $w(t) = (1/\sqrt{2\pi\hat{\sigma}_n^2})e^{-t^2/(2\hat{\sigma}_n^2)}$.

$$I_n = \int_{-\infty}^{\infty} |\hat{\phi}_S(t) - \hat{\phi}_0(t)|^2 \, w(t)dt$$

$$= n^{-2}\sum_{j=1}^{n}\sum_{k=1}^{n}\exp\{-\tfrac{1}{2}(X_j - X_k)^2/\hat{\sigma}_n^2\} - 2n^{-1}\sum_{j=1}^{n}\exp\{-\tfrac{1}{2}(X_j - \bar{X}_n)^2/\hat{\sigma}_n^2\}$$

$$+ 1/\sqrt{3}. \qquad (9)$$

Henze (1990) derived a large sample approximation for $T_n = nI_n$ and used Pearson curves to approximate the distribution. A simulation study was conducted to calculate the $1-\alpha$ percentiles of $T_n$, $\alpha = 0.05$, based on $m = 200000$ simulated values of $T_n$. It was found that even with this large sample there is still variation in the 4$^{th}$ decimal and the first 3 decimals where used in the simulation study to estimate the power of the test.

Table 1. Simulated percentiles to test for normality using the Epps-Pulley test at the 5% level. Calculated from $m = 200000$ simulated samples of size $n$

| $n$ | .95 percentile | mean | variance |
|---|---|---|---|
| 50 | 0.370 | 0.1303 | 0.0148 |
| 100 | 0.373 | 0.1321 | 0.0150 |
| 250 | 0.375 | 0.1338 | 0.0151 |
| 500 | 0.377 | 0.1338 | 0.0152 |
| 750 | 0.377 | 0.1334 | 0.0151 |
| 1000 | 0.377 | 0.1336 | 0.0151 |

**Table 1** Simulated percentiles to test for normality using the Epps-Pulley test at the 5% level. Calculated from $m = 200000$ simulated samples of size $n$.

Henze (1990) used the value 1 instead of $\hat{\sigma}_n^2$ and found for example for $n = 100$, that the .95$^{th}$ percentile is 0.376 which is approximately equal to 0.373 found in this study. They calculated the asymptotic expected value of $T_n$ as 0.13397 and variance 0.015236.

# 3. SIMULATION STUDY

The paper of Yap and Sim (2011) is used as a guideline to decide which tests to include. The proposed test will be denoted by ECFT and EP denotes the Epps-Pulley test in the tables. The power of the test will be compared against several tests for normality:

- The Lilliefors test (LL), Lilliefors (1967) which is a slight modification of the Kolmogorov-Smirnov test for where parameters are estimated.

- The Jarque-Bera test (JB), Jarque and Bera (1987), where the skewness and kurtosis is combined to form a test statistics.

- The Shapiro-Wilks test (SW), Shapiro and Wilk (1965). This test makes use of properties of order statistics and were later developed to be used for large samples too by Royston (1992).

- The Anderson-Darling test (AD), Anderson and Darling .

- The D'Agostino and Pearson test (DP), D'Agostino and Pearson (1973). This statistic combines the skewness and kurtosis to check for deviations from normality.

Samples are generated from a few symmetric unimodal symmetric distributions with sizes $n = 50, 100, 250, 500, 750, 1000$. The proportion rejections are reported based on $m$=5000 repetitions. The test are conducted at the 5% level and for the ecf, the normal approximation is used. Since the sample sizes are large, non-normality with respect to multi-modal and skewed distributions can easily be picked up by using graphical methods.

The following symmetric distributions are considered, uniform on the interval [0,1], the logistic distribution with mean zero the standard t-distribution and the Laplace distribution with mean zero and scale parameter one. The standard $t$-distribution with 4, 10 and 15 degrees of freedom. Skewed distributions and multimodal distributions would not be investigated, since in large samples such samples can be already excluded with certainty as being not from a normal distribution by looking at the histograms.

All the tests performed for the ecf test were conducted using the normal approximation, but percentiles can also easily be simulated. Simulated estimates of the Type I error for $n = 30, 50, 100, 250, 500, 750, 1000$, are given in Table 1 based on $m = 5000$ simulated samples. The simulated samples are standard normally distributed and studentized to calculate the Type I error.

Table 2  Simulated percentiles to test for normality at the 5% level. Calculated from $m = 5000$ simulated samples of size $n$ each in the point $t=1$.

| | **Type I error** | | | | | | |
|---|---|---|---|---|---|---|---|
| $n$ | **ECFT** | **EP** | **LL** | **JB** | **SW** | **AD** | **DP** |
| 50 | 0.0406 | 0.0476 | 0.0544 | 0.0484 | 0.0478 | 0.0506 | 0.0568 |
| 100 | 0.0460 | 0.0508 | 0.0468 | 0.0524 | 0.0460 | 0.0496 | 0.0602 |
| 250 | 0.0470 | 0.0536 | 0.0574 | 0.0546 | 0.0454 | 0.0526 | 0.0582 |
| 500 | 0.0470 | 0.0556 | 0.0526 | 0.0526 | 0.0442 | 0.0546 | 0.0510 |
| 750 | 0.0512 | 0.0528 | 0.0480 | 0.0522 | 0.0432 | 0.0532 | 0.0520 |

| | 1000 | 0.0478 | 0.0496 | 0.0486 | 0.0482 | 0.0466 | 0.0456 | 0.0500 |

In Table 2 the rejection rates, when testing at the 5% level and symmetric distributions, are shown for various sample sizes based on 10000 samples each time.

In table 1 the t-distribution where not all moments exist is considered. The JB, DP and ECFT tests performs best, and in large samples the ECFT test performs the best.

Table 3. Simulated power of normality tests. Rejection proportions when testing for normality at the 5% level.

| | n | ECFT | EP | LL | JB | SW | AD | DP |
|---|---|---|---|---|---|---|---|---|
| t(4) | 50 | 0.5474 | 0.4296 | 0.2892 | 0.5206 | 0.4504 | 0.4034 | 0.4912 |
| | 100 | 0.8062 | 0.6684 | 0.4742 | 0.7692 | 0.7002 | 0.6388 | 0.7256 |
| | 250 | 0.9892 | 0.9574 | 0.8394 | 0.9790 | 0.9636 | 0.9410 | 0.9654 |
| | 500 | 1.0000 | 0.9998 | 0.9866 | 0.9996 | 0.9994 | 0.9992 | 0.9994 |
| | 750 | 1.0000 | 1.0000 | 0.9992 | 1.0000 | 1.0000 | 1.0000 | 1.0000 |
| | 1000 | 1.0000 | 1.0000 | 0.9998 | 1.0000 | 1.0000 | 1.0000 | 1.0000 |
| t(10) | 50 | 0.2078 | 0.1350 | 0.0888 | 0.2026 | 0.1558 | 0.1218 | 0.1896 |
| | 100 | 0.3238 | 0.1748 | 0.1034 | 0.3010 | 0.2142 | 0.1588 | 0.2592 |
| | 250 | 0.5692 | 0.3216 | 0.1652 | 0.5296 | 0.4126 | 0.2774 | 0.4626 |
| | 500 | 0.8062 | 0.5298 | 0.2622 | 0.7544 | 0.6376 | 0.4676 | 0.6934 |
| | 750 | 0.9174 | 0.7138 | 0.4110 | 0.8816 | 0.7970 | 0.6544 | 0.8450 |
| | 1000 | 0.9648 | 0.8296 | 0.5112 | 0.9422 | 0.8862 | 0.7756 | 0.9208 |
| t(15) | 50 | 0.1368 | 0.0952 | 0.0710 | 0.1420 | 0.1080 | 0.0864 | 0.1360 |
| | 100 | 0.2100 | 0.1156 | 0.0736 | 0.2066 | 0.1488 | 0.1024 | 0.1822 |
| | 250 | 0.3462 | 0.1804 | 0.0918 | 0.3210 | 0.2252 | 0.1524 | 0.2728 |

|   | 500  | 0.5144 | 0.2540 | 0.1292 | 0.4800 | 0.3550 | 0.2190 | 0.4226 |
|---|------|--------|--------|--------|--------|--------|--------|--------|
|   | 750  | 0.6608 | 0.3542 | 0.1736 | 0.6132 | 0.4670 | 0.3072 | 0.5482 |
|   | 1000 | 0.7744 | 0.4724 | 0.2208 | 0.7148 | 0.5888 | 0.4090 | 0.6646 |

Fig. 4. Plot of the three best performing tests with respect to power, testing for normality, data t-distributed with 10 degrees of freedom. Solid line, ecf test, dashed line JB test and dash-dot the DP test.

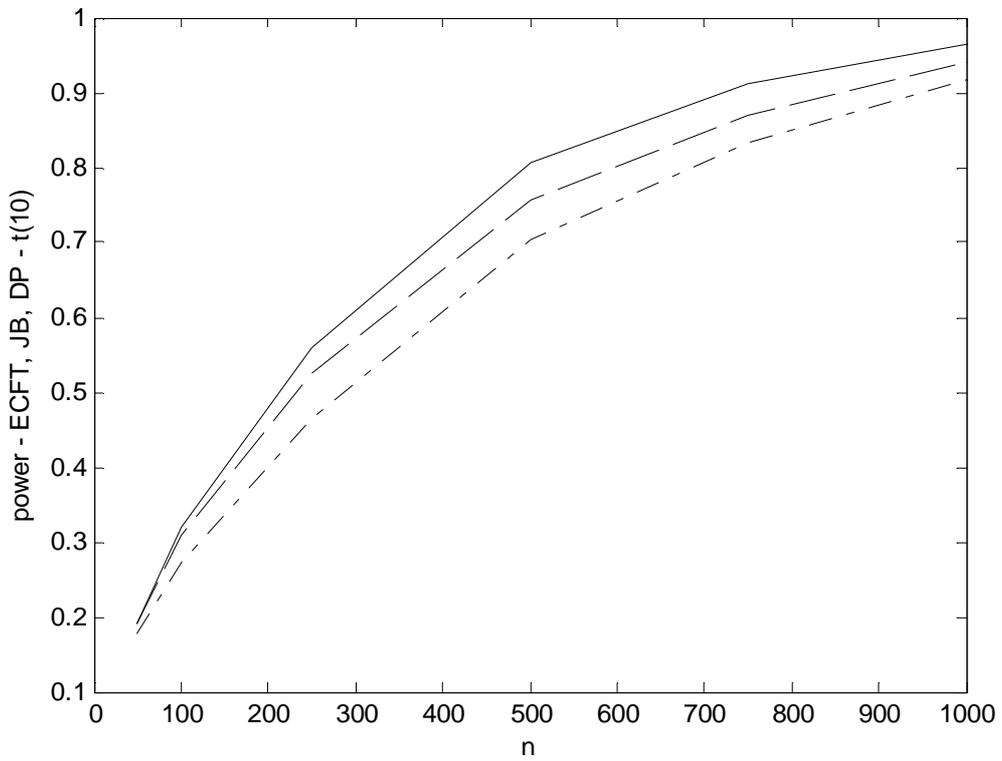

Table 4. Simulated power of normality tests. Rejection proportions when testing for normality at the 5% level.

|        | n    | ECFT   | EP     | LL     | JB     | SW     | AD     | DP     |
|--------|------|--------|--------|--------|--------|--------|--------|--------|
| U(0,1) | 50   | 0.1330 | 0.4654 | 0.2560 | 0.0078 | 0.5748 | 0.5640 | 0.7986 |
|        | 100  | 0.9496 | 0.9228 | 0.5860 | 0.7396 | 0.9844 | 0.9486 | 0.9962 |
|        | 250  | 1.0000 | 1.0000 | 0.9862 | 1.0000 | 1.0000 | 1.0000 | 1.0000 |
|        | 500  | 1.0000 | 1.0000 | 0.9862 | 1.0000 | 1.0000 | 1.0000 | 1.0000 |
| Laplace| 50   | 0.6100 | 0.5356 | 0.4380 | 0.5572 | 0.5126 | 0.5430 | 0.5130 |
|        | 100  | 0.8666 | 0.8148 | 0.7026 | 0.8008 | 0.7804 | 0.8276 | 0.7326 |
|        | 250  | 0.9974 | 0.9958 | 0.9802 | 0.9876 | 0.9896 | 0.9964 | 0.9756 |
|        | 500  | 1.0000 | 1.0000 | 1.0000 | 1.0000 | 1.0000 | 1.0000 | 1.0000 |
| Logistic| 50  | 0.2694 | 0.1798 | 0.1188 | 0.2598 | 0.1946 | 0.1612 | 0.2406 |
|        | 100  | 0.4334 | 0.2700 | 0.1550 | 0.3960 | 0.2990 | 0.2344 | 0.3472 |
|        | 250  | 0.7316 | 0.5122 | 0.2838 | 0.6744 | 0.5576 | 0.4630 | 0.6024 |
|        | 500  | 0.9344 | 0.7960 | 0.5164 | 0.8876 | 0.8264 | 0.7514 | 0.8494 |
|        | 750  | 0.9838 | 0.9318 | 0.7086 | 0.9702 | 0.9410 | 0.9112 | 0.9534 |
|        | 1000 | 0.9958 | 0.9756 | 0.8246 | 0.9902 | 0.9792 | 0.9614 | 0.9846 |

It can be seen the ECFT outperforms the other tests with respect to power in large samples, especially when testing data from a logistic distribution.

Samples were simulated from a mixture of two normal distributions, with a proportion $\alpha$ from a standard normal and a proportion $1-\alpha$ from a normal with variance $\sigma^2$. This can also be considered as a contaminated distribution. The results are shown in Table 4.

The proposed test yielded good results.

Table 5. Simulated power of normality tests. Rejection proportions when testing for normality at the 5% level. Mixture of two normal distributions (contaminated data).

| $(\sigma,\alpha)$ | n | ECFT | EP | LL | JB | SW | AD | DP |
|---|---|---|---|---|---|---|---|---|
| (2.0,0.2) | 50 | 0.3720 | 0.2384 | 0.1362 | 0.3648 | 0.2746 | 0.2118 | 0.3346 |
| | 100 | 0.5838 | 0.3700 | 0.1884 | 0.5512 | 0.4426 | 0.3244 | 0.4976 |
| | 250 | 0.8804 | 0.6776 | 0.3696 | 0.8482 | 0.7640 | 0.6134 | 0.7928 |
| | 500 | 0.9838 | 0.9112 | 0.6342 | 0.9748 | 0.9516 | 0.8784 | 0.9610 |
| | 750 | 0.9992 | 0.9844 | 0.8258 | 0.9992 | 0.9960 | 0.9756 | 0.9978 |
| | 1000 | 0.9998 | 0.9966 | 0.9132 | 0.9998 | 0.9982 | 0.9934 | 0.9994 |
| (0.5,0.2) | 50 | 0.1122 | 0.0964 | 0.0764 | 0.1078 | 0.0888 | 0.0946 | 0.1000 |
| | 100 | 0.1568 | 0.1114 | 0.0902 | 0.1400 | 0.0922 | 0.1088 | 0.1164 |
| | 250 | 0.2786 | 0.1922 | 0.1494 | 0.2142 | 0.1494 | 0.1984 | 0.1710 |
| | 500 | 0.4580 | 0.3486 | 0.2532 | 0.3390 | 0.2490 | 0.3562 | 0.2676 |
| | 750 | 0.5796 | 0.4766 | 0.3418 | 0.4222 | 0.3464 | 0.4956 | 0.3486 |
| | 1000 | 0.7110 | 0.6284 | 0.4668 | 0.5420 | 0.4822 | 0.6352 | 0.4710 |
| (2.0,0.5) | 50 | 0.3012 | 0.2106 | 0.1382 | 0.2652 | 0.1990 | 0.1982 | 0.2368 |
| | 100 | 0.4822 | 0.3486 | 0.2200 | 0.4088 | 0.3198 | 0.3288 | 0.3416 |
| | 250 | 0.8250 | 0.6972 | 0.4568 | 0.7158 | 0.6316 | 0.6642 | 0.6172 |
| | 500 | 0.9800 | 0.9484 | 0.7714 | 0.9406 | 0.9168 | 0.9356 | 0.8994 |
| | 750 | 0.9984 | 0.9938 | 0.9290 | 0.9924 | 0.9870 | 0.9914 | 0.9836 |
| | 1000 | 0.9998 | 0.9992 | 0.9782 | 0.9976 | 0.9970 | 0.9986 | 0.9962 |

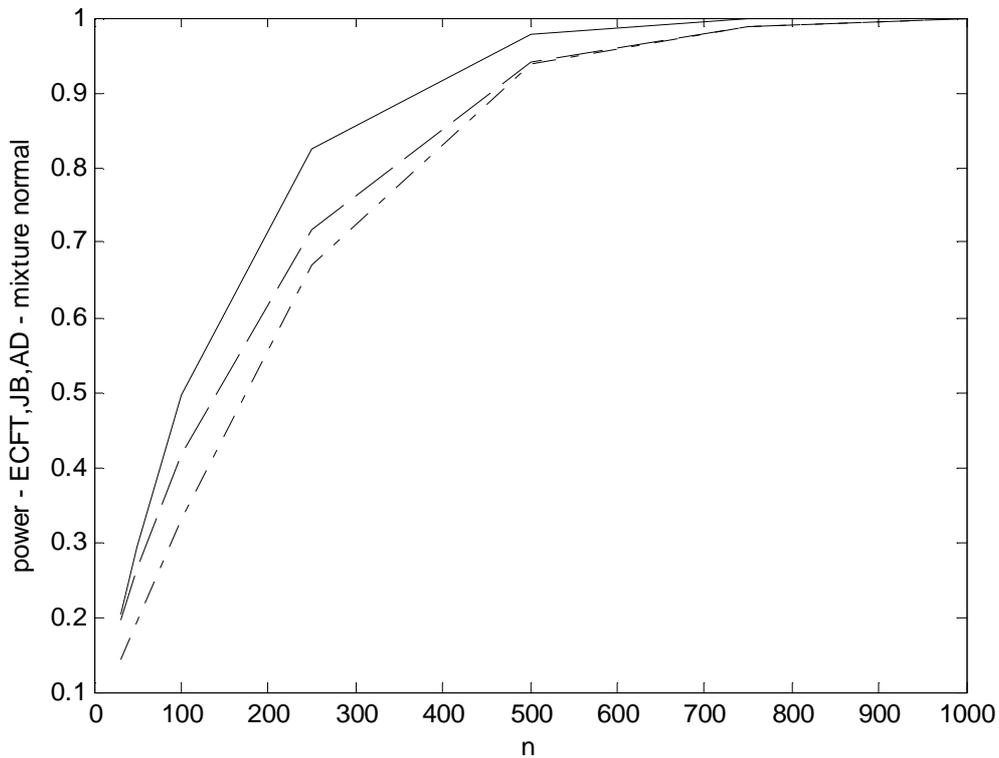

Fig. 5. Plot of the three best performing tests with respect to power, testing for normality, data mixture of normal distributions with two components. Mixture .5N(0,1)+0.5N(0,4). Solid line, ecf test, dashed line JB test and dash-dot the AD test.

## 4. CONCLUSIONS

The proposed test performs better with respect to power in large samples than the other tests for normality for the distributions considered in the simulation study . In small samples of say less than $n=50$, it was found that the test of D'Agostino and Pearson (1973) was often either the best performing or close to the best performing test.

In practice one would not test data from a skewed distribution for normality in large samples. The simple normal test approximation will perform better, the larger the sample is. The asymptotic normality and variance properties, which is of a very simple form, can be used in large samples. This test can be recommended as probably the test of choice in terms of power and easy of application in large samples and shows that the empirical characteristic function has the potential to outperform the usual frequentist methods.

## References

ANDERSON, T.W., DARLING, D.W. (1954). A test of goodness of fit. *J. Amer. Statist. Assoc.,* 49 (268), 765 – 769.

BARINGHAUS, L., HENZE, N. (1988). A consistent test for multivariate normality based on the empirical characteristic function. Metrika 35 (1): 339–348.

CRAMÉR, H. (1946). Mathematical Methods of Statistics. NJ: Princeton University Press, Princeton, NJ.

CSÖRGŐ, S. (1986). Testing for normality in arbitrary dimension. *Annals of Statistics*, 14 (2), 708 – 723.

D'AGOSTINO, R., PEAESON, R. (1973). Testing for departures from normality. *Biometrika*, 60, 613 - 622.

EPPS, T.W., PULLEY, L.B. (1983). A test for normality based on the empirical characteristic function. *Biometrika*, 70, 723 -726.

FEUERVERGER, A., MUREIKA, R.A. (1977). The Empirical Characteristic Function and Its Applications. *The Annals of Statistics*, 5 (1), 88 – 97.

HENZE, N. (1990), An Approximation to the Limit Distribution of the Epps-Pulley Test Statistic for Normality. *Metrika*, 37, 7 – 18.

JARQUE, C.M., BERA, A.K. (1987). A test for normality of observations and regression residuals. *Int. Stat. Rev.* , 55 (2), 163 – 172.

LILLIEFORS, H.W. (1967). On the Kolmogorov-Smirnov test for normality with mean and variance unknown. *J. American. Statist. Assoc.,* 62, 399 – 402.

LUKACS, E. (1972). A Survey of the Theory of Characteristic Functions. *Advances in Applied Probability*, 4 (1), 1 – 38.

MEINTANIS, S. G. (2016). A review of testing procedures based on the empirical characteristic function. *South African Statist. J.* , 50, 1 – 14.

MUROTA, K., TAKEUCHI, K. (1981). The studentized empirical characteristic function and its application to test for the shape of distribution. *Biometrika*; 68, 55 -65.

ROMÃO, X, DELGADO, R , COSTA, A. (2010). An empirical power comparison of univariate goodness-of-fit tests for normality. J. of Statistical Computation and Simulation.; 80:5, 545 – 591.

ROYSTON, J.P. (1992). Approximating the Shapiro-Wilk W-test for non-normality. *Stat. Comput.,* 2, 117 – 119.

SHAPIRO, S.S., WILK, M.B. (1965). An analysis of variance test for normality (complete samples). Biometrika 52, 591 -611.

SWANEPOEL, J.W.H., ALLISON, J. (2016). Comments: A review of testing procedures based on the empirical characteristic function. *South African Statist. J.* , 50, 1 – 14.

TAUFER, E. (2016). Comments: A review of testing procedures based on the empirical characteristic function. *South African Statist. J.* , 50, 1 – 14.

USHAKOV, N.G. (1999). Selected Topics in Characteristic Functions. VSP, Utrecht.

YAP, B.W., SIM, C.H. (2011). Comparisons of various types of normality tests. *J. of Statistical Computation and Simulation*, 81 (12), 2141 - 2155.